\documentclass[10pt,english,aps,prd,superscriptaddress,nofootinbib,notitlepage,preprintnumbers]{revtex4-1}
\usepackage[T1]{fontenc}
\usepackage[latin9]{inputenc}
\usepackage{amsmath}
\usepackage{graphicx}

\makeatletter
\def\Mpl{M_{\rm P}}

\usepackage{epstopdf}

\usepackage{babel}

\@ifundefined{showcaptionsetup}{}{%
 \PassOptionsToPackage{caption=false}{subfig}}
\usepackage{subfig}
\makeatother

\begin{document}
\preprint{YITP-20-43}
\title{Anisotropic instability in a higher order gravity theory}
\author{Masroor C.~Pookkillath}
\email{masroor.cp@yukawa.kyoto-u.ac.jp}

\affiliation{Center for Gravitational Physics, Yukawa Institute for Theoretical
Physics, Kyoto University, 606-8502, Kyoto, Japan}
\author{Antonio De~Felice}
\email{antonio.defelice@yukawa.kyoto-u.ac.jp}

\affiliation{Center for Gravitational Physics, Yukawa Institute for Theoretical
Physics, Kyoto University, 606-8502, Kyoto, Japan}
\author{Alexei A.~Starobinsky}
\email{alstar@landau.ac.ru}

\affiliation{L.\ D.\ Landau Institute for Theoretical Physics, Moscow 119334,
Russian Federation}
\affiliation{Kazan Federal University, Kazan 420008, Republic of Tatarstan, Russian
	Federation}
\date{\today}
\begin{abstract}
  We study a metric cubic gravity theory considering odd-parity modes of 
  linear inhomogeneous perturbations on a spatially homogeneous 
  Bianchi type I manifold close to the isotropic de Sitter spacetime. We 
  show that in the regime of small anisotropy, the theory possesses new 
  degrees of freedom compared to General Relativity, whose kinetic energy 
  vanishes in the limit of exact isotropy. From the mass dispersion relation 
  we show that such theory always possesses at least one ghost mode as 
  well as a very short-time-scale (compared to the Hubble time) classical 
  tachyonic (or ghost-tachyonic) instability. In order to confirm our analytic 
  analysis, we also solve the equations of motion numerically and we find 
  that this instability is developed well before a single e-fold of the scale 
  factor. This shows that this gravity theory, as it is, cannot be used to 
  construct viable cosmological models.
\end{abstract}
\maketitle

\section{Introduction}

Modified gravity models have been introduced in theoretical physics
for different aims. General covariance allows for an infinite number
of geometrical Lagrangian densities built only out of the Riemann tensor, 
its covariant derivatives and its contractions. The first and simplest example 
is for sure the Einstein-Hilbert action, which was giving the original
1916 Einstein equations~\cite{1915SPAW.......844E,1916AnP...354..769E},
and up to now, with the addition of a cosmological constant $\Lambda$ 
introduced by Einstein, too~\cite{Einstein:1917ce}, it is considered to be the theory of 
gravitational interactions valid both at small and large (cosmological) scales.

However, modifications of General Relativity (GR) beyond the Ricci scalar $R$ and 
$\Lambda$, were studied in the context of efforts to obtain a renormalizable 
theory of gravity~\cite{Stelle:1976gc}. But probably the biggest success in 
application to observable effects was achieved by the introduction of such a 
modification in cosmology in the context of 
inflation~\cite{Starobinsky:1980te,Mukhanov:1981xt,Starobinsky:1983zz}. 
Since then, modified gravity theories were considered not only as purely 
theoretical constructions, but they have become the basis for building 
falsifiable models able to describe high curvature regimes in gravity and 
cosmology~\cite{Horava:2008ih,Horava:2009uw,Oliva:2010eb,Myers:2010ru,Hennigar:2017ego,Cisterna:2018tgx}.

More recently, after the discovery of the late time accelerated expansion of 
our Universe, purely geometrical modified gravity theories like the $f(R)$ 
gravity~\cite{DeFelice:2010aj} have been also considered as alternative to scalar 
field (quintessence) models in GR in order to describe present dark energy
in the Universe not by $\Lambda$, but through infra-red modifications of GR, 
see the review~\cite{Clifton:2011jh}. This has also shed new light on other, 
more general modifications of gravity like scalar-tensor 
theories~\cite{Fujii:2003pa,Kobayashi:2019hrl} including $f(R)$ gravity as 
a particular case, vector-tensor theories~\cite{Heisenberg:2017mzp,Heisenberg:2016eld},
massive gravity~\cite{deRham:2010ik,deRham:2010kj,DeFelice:2015hla},
bi-gravity \cite{Rosen:1975kk} and so on.

All these modifications of gravity tend to share, as a common feature,
the property of appearance of new degrees of freedom besides the standard massless
gravitational waves, even in the absence of matter. These new degrees
of freedom in general lead to cosmological low-curvature phenomenology 
completely different from the standard $\Lambda$-CDM model, unless some 
mechanism exists to effectively screen them like the chameleon 
mechanism~\cite{Khoury:2003aq}. In addition, for all purely geometrical
modifications of GR without torsion and non-metricity known by now, 
apart from $f(R)$ gravity satisfying the conditions $f'(R)>0,~f''(R)>0$, 
these degrees of freedom appear to be tachyons or ghosts, or even 
ghost-tachyons.

Among these modifications of gravity, there recently appeared a model
which imposes, as a defining condition, that it leads to equations of motion 
which are only of the second order on maximally symmetric 
spacetimes~\cite{Bueno:2016xff,Arciniega:2018tnn}. Thus, the theory is 
required to possess the same number of degrees of freedom as GR on these 
spacetimes. This condition seem not to agree with the Lovelock condition 
which requires second order differential equations on \emph{any} 
background~\cite{Lovelock:1971yv}. As a result, several authors found 
Lagrangians not reducing to the Lovelock result in four space-time 
dimensions. In fact, at least a Lagrangian cubic in the Riemann tensor (and 
its contractions), known as Einsteinian cubic gravity (ECG) has been 
introduced in~\cite{Bueno:2016xff,Arciniega:2018fxj}. This model was later generalized to all orders in the Riemann tensor in~\cite{Arciniega:2018tnn}

However, since ECG does not satisfy the Lovelock theorem, one should 
expect that it generically possesses other propagating modes. On a FLRW 
spacetime without matter and with second order equations of motion for 
the scale factor and for perturbations, these modes can be tensor 
perturbations (massive gravitational waves) only. This fact could be 
problematic, but if the mass of these extra degrees 
of freedom is large enough, then the model would still be viable as an 
effective low-energy theory. If so, the above mentioned defining 
prescription on maximally symmetric spacetimes could be enough to 
ensure good behavior of higher order gravity theories. That is why in this 
paper, we will try to understand what happens to the extra degrees of 
freedom, and in particular, we will address the issue whether these extra 
degrees of freedom make the maximally symmetric de Sitter solution 
unstable or not.

For this purpose, we will find it convenient to study ECG solutions for 
small inhomogeneous perturbations on a plane-symmetric spatially 
homogeneous Bianchi type I spacetime. We consider this particular 
manifold because, as we will show later on, it possesses a smooth limit 
to an isotropic and homogeneous FLRW background. The Bianchi-I 
spacetime itself can be thought of as a strong tensor perturbation 
(gravitational wave) with the infinite wavelength superimposed on a 
FLRW background. Therefore, if we use solutions for perturbations on it 
to study what happens to the extra degrees of freedom as the Bianchi-I 
metric becomes more and more isotropic, this consideration will be 
effectively beyond the linear order with respect to the limiting FLRW 
background. To arrive such a goal, we will find it sufficient to study only 
the odd-parity-modes subset of all perturbation variables. We find that, 
on a general Bianchi type I manifold, three odd-parity modes propagate 
(instead of the single one present in GR). This fact is in agreement with 
the Lovelock theorem. In other words, solutions of ECG for perturbations 
on a generic background do not behave as in GR. In fact, we find that 
for any Bianchi I type background solution in ECG, at least one of its 
three perturbation modes is always a ghost.

However, these anisotropic degrees of freedom are not present on a 
FLRW background. Therefore, we want to know how the ECG theory 
behaves in the FLRW limit. We show the existence of such a FLRW limit, 
i.e.\ the existence of a Bianchi type I background solution of the ECG 
equations of motion which evolves in time more and more towards a 
FLRW isotropic solution. We will call it the isotropic limit of Bianchi-I 
solutions. This solution is important as it will show what happens to the 
three anisotropic modes as the background becomes more and more 
isotropic.

In this isotropic limit, we then study the no-ghost conditions for
the three modes, and find that at all times either one or two modes
are ghosts. We then proceed to study the dispersion relations for
all the modes in this isotropic limit. We find that at the leading
order, the dynamics of the ghost(s) decouple from the other modes,
and one (of two) modes become either tachyonic or ghost-tachyonic
(we will see more clearly what we mean by this later on). Anyhow,
in both these cases, a strong \emph{classical} instability arises,
which exponentially grows in a short time (much less than in a single
e-fold).

For this reason, we believe that the maximally symmetric FLRW 
spacetime -- the de Sitter one -- cannot be considered as a stable, 
ghost-free solution of the equations of motion. Rather, as long as the 
background is not exactly de Sitter, even if the would-be FLRW 
background equations of motion are stable, tensor perturbations
(at least their odd-parity modes) will have extra and/or ghost degrees 
of freedom which will tend to grow exponentially. Therefore, as long 
as one does not find a cure to such a behavior, the ECG theory could 
not lead to a viable cosmological model.

This paper is organized as follows. In section \ref{Lagrangian} we
introduce the ECG Lagrangian and discuss small inhomogeneous odd-parity perturbations of 
a homogeneous Bianchi-I metric. In section \ref{back_FLRWlimit}, we establish the existence
of an isotropic limit solution to the Bianchi-I metric. Then, we expand the
Lagrangian up to the second order in odd-parity modes in section \ref{dof}.
Introducing a new Lagrangian multiplier, we find that there exist three
degrees of freedom for perturbations. After diagonalizing the kinetic matrix, 
we make a canonical field redefinition in the section \ref{mdr} to study
the mass dispersion relation. In this section we show that there always
exists at least one ghost and tachyonic or ghost-tachyonic instability.
In section, \ref{num} we solve the equation of motion numerically
with respect to the number of scale factor e-folds $N$, and show that the instability
is developed much before even single e-fold number. Finally in section
\ref{conclusion}, we present our concluding remarks.

\section{The Lagrangian density\label{Lagrangian}}

The Lagrangian density of the ECG contains the following linear combination 
of cubic in the Riemann and Ricci tensors terms~\cite{Arciniega:2018fxj} 
\begin{equation}
\mathcal{L}\equiv12\mathcal{L}_{1}+\mathcal{L}_{2}-8\mathcal{L}_{3}+2\mathcal{L}_{4}+4\mathcal{L}_{5}+8\mathcal{L}_{6}-4\mathcal{L}_{7}\,,
\end{equation}
where 
\begin{eqnarray}
\mathcal{L}_{1} & = & R^{\alpha}{}_{\beta}{}^{\gamma}{}_{\delta}\,R^{\beta}{}_{\mu}{}^{\delta}{}_{\nu}\,R^{\mu}{}_{\alpha}{}^{\nu}{}_{\gamma}\,,\\
\mathcal{L}_{2} & = & R^{\alpha\beta}{}_{\gamma\delta}\,R^{\gamma\delta}{}_{\mu\nu}\,R^{\mu\nu}{}_{\alpha\beta}\,,\\
\mathcal{L}_{3} & = & R_{\alpha\beta\gamma\delta}\,R^{\alpha\beta\gamma}{}_{\mu}\,R^{\delta\mu}\,,\\
\mathcal{L}_{4} & = & R_{\alpha\beta\gamma\delta}\,R^{\alpha\beta\gamma\delta}\,R\,,\\
\mathcal{L}_{5} & = & R_{\alpha\beta\gamma\delta}\,R^{\alpha\gamma}\,R^{\beta\delta}\,,\\
\mathcal{L}_{6} & = & R^{\alpha}{}_{\beta}\,R^{\beta}{}_{\delta}\,R^{\delta}{}_{\alpha}\,,\\
\mathcal{L}_{7} & = & R^{\alpha}{}_{\beta}\,R^{\beta}{}_{\alpha}\,R\,,
\end{eqnarray}
and total gravitational action reads as 
\begin{equation}
S=\int d^{4}x\sqrt{-g}\left[\frac{\Mpl^{2}}{2}\,(R-2\Lambda)+\frac{\beta}{\Mpl^{2}}\,\mathcal{L}\right].
\label{action}
\end{equation}

However, the scalar ${\mathcal L}$, not belonging to any of the
Lovelock scalars, leads to non-trivial contributions in four
dimensions, and in turn, this new theory of gravity is then, if seen
from the Lovelock theorem point of view, necessarily of higher
order. Nonetheless, in vacuum and on a spatially flat FLRW 
background, it is rather easy to show that the following background 
equations of motion hold
\begin{eqnarray}
\frac{\dot{a}^{2}}{a^{2}}+\frac{16\beta}{\Mpl^{4}}\,\frac{\dot{a}^{6}}{a^{6}}-\frac{\Lambda}{3} & = & 0\,,\label{eq:1stFRIED}\\
\left(\frac{\ddot{a}}{a}-\frac{\dot{a}^{2}}{a^{2}}\right)\left(\Mpl^{4}+48\beta\,\frac{\dot{a}^{4}}{a^{4}}\right) & = & 0\,.\label{eq:2nd FRIED}
\end{eqnarray}
These equations of motion imply a de Sitter solution $a\propto e^{H_{0}t}$,
where $H_{0}=\frac{\dot{a}}{a}$ and the constant value of $H_{0}$ is
related to the value of the bare cosmological constant $\Lambda$. On top of
that, considering linear perturbation theory, we find that only
the two different polarization of the tensor modes, 
i.e.\ $h_{ij}=\sum_{\lambda={+},{\times}}H_{\lambda}\epsilon_{ij}^{\lambda}$
do propagate, and their reduced action can be written as 
\begin{equation}
S=\frac{\Mpl^{4}+48\beta H_{0}^{4}}{8\Mpl^{2}}\sum_{\lambda}\int dt\,d^{3}x\,a^{3}\left[\dot{H}_{\lambda}^{2}-\frac{1}{a^{2}}\,(\partial_{i}H_{\lambda})^{2}\right].
\end{equation}
Indeed, this theory satisfies the defining conditions that on
maximally symmetric spacetimes, the equations of motion are of second
order and only tensor modes propagate, as in GR.

This result also shows that, as long as $\Mpl^{4}+48\beta H_{0}^{4}>0$, 
the tensor modes are well behaved on the de Sitter solution. This necessary 
condition automatically excludes the particular value of $H_{0}$ which 
would make the second factor in Eq.\ (\ref{eq:2nd FRIED}) vanish. 
Furthermore, this condition together with Eq.\ (\ref{eq:1stFRIED}) imply 
that the de Sitter solution exists provided $ \Lambda > 2H_0^2 > 0 $. Therefore, 
there are no stable de Sitter solutions without a bare positive cosmological 
constant in this gravity theory. The condition $\Lambda>2H_0^2$ is very strong, 
as this fact indicates that in general there is no mechanism to end inflation 
within the theory itself, at least at the classical level. This immediately puts 
a very serious obstacle to the construction of viable cosmological models in 
this theory. Still it is possible to avoid it, i.e., like it was done 
in ~\cite{Starobinsky:1980te}, by waiving the requirement that the equation 
of motion for the scale factor should be the second order (but still keeping 
second order equations for perturbations!) and by adding the $R^2$ term to 
the total Lagrangian density in the rhs of (\ref{action}). However, as we shall 
see later on, the real problem of this theory is that this de Sitter solution is 
unstable due to the presence of perturbation instabilities.

\subsection{The metric}

Let us consider a homogeneous plane-symmetric Bianchi type I background 
\begin{equation}
ds^{2}=-dt^{2}+a^{2}\,dx^{2}+b^{2}\,\delta_{ij}dy^{i}dy^{j}\,,
\end{equation}
and let us focus on odd-parity modes of perturbations upon it (dubbed odd modes below for brevity):
\begin{eqnarray}
ds^{2} & = & -dt^{2}+a^{2}\,dx^{2}+b^{2}\,\delta_{ij}\,dy^{i}dy^{j}\nonumber \\
 &  & {}+2bV_{i}\,dt\,dy^{i}+2\,ab\,(\partial_{x}W_{i})\,dx\,dy^{i}+\frac{b}{2}\,(\partial_{i}Z_{j}+\partial_{j}Z_{i})\,dy^{i}dy^{j}\,.
\end{eqnarray}
For an odd modes gauge transformation, we get 
\begin{equation}
\xi^{\alpha}=(0,0,\xi^{i})\,,
\end{equation}
where 
\begin{equation}
\delta^{ij}\partial_{j}V_{i}=0=\delta^{ij}\partial_{j}W_{i}=0=\delta^{ij}\partial_{j}Z_{i}=0=\partial_{i}\xi^{i}\,,
\end{equation}
because the metric in the $y$-subspace is Euclidean. Then we can
consider a decomposition for odd modes as follows 
\begin{eqnarray}
V_{i} & = & \frac{V(t,x)}{b}\,\epsilon_{ij}\delta^{jk}\partial_{k}Y(y)\,,\\
W_{i} & = & \frac{W(t,x)}{b}\,\epsilon_{ij}\delta^{jk}\partial_{k}Y(y)\,,\\
Z_{i} & = & \frac{Z(t,x)}{b}\,\epsilon_{ij}\delta^{jk}\partial_{k}Y(y)\,,\\
\xi^{i} & = & \frac{\xi_{V}(t,x)}{b}\,\delta^{il}\,\epsilon_{lj}\delta^{jk}\partial_{k}Y(y)\,,
\end{eqnarray}
where $Y$ satisfies the equation $\delta^{ij}\partial_{i}\partial_{j}Y=-q^{2}\,Y$,
and $\epsilon_{12}=1=-\epsilon_{21}$. In fact, we find 
\begin{equation}
\delta^{ij}\partial_{j}V_{i}=V(t,x)\,\epsilon_{il}\delta^{lk}\delta^{ij}\partial_{j}\partial_{k}Y(y)=V(t,x)\,\epsilon^{jk}\partial_{j}\partial_{k}Y(y)=0\,,
\end{equation}
automatically.

For an odd modes gauge transformation, we have 
\begin{eqnarray}
\Delta\delta g_{ij} & = & -\xi_{i;j}-\xi_{j;i}=-(\xi_{i,j}-\Gamma^{\alpha}{}_{ij}\,\xi_{\alpha})-(\xi_{j,i}-\Gamma^{\alpha}{}_{ji}\,\xi_{\alpha})\nonumber \\
 & = & -\xi_{i,j}-\xi_{j,i}+2\Gamma^{l}{}_{ij}\,\xi_{l}=-\xi_{i,j}-\xi_{j,i}\,.
\end{eqnarray}
This leads to 
\begin{equation}
\Delta Z=-2\xi_{V}/b\,.
\end{equation}

So we set the odd-mode flat gauge for which 
\begin{equation}
Z=0\,,
\end{equation}
or

\begin{eqnarray}
ds^{2} & = & -dt^{2}+a^{2}\,dx^{2}+b^{2}\,\delta_{ij}\,dy^{i}dy^{j}\nonumber \\
 &  & {}+2V(t,x)\,\epsilon_{ij}\delta^{jk}\partial_{k}Y(y)\,dt\,dy^{i}+2\,a\,\partial_{x}W(t,x)\,\epsilon_{ij}\delta^{jk}\partial_{k}Y(y)\,dx\,dy^{i}\,.
\end{eqnarray}
In the limit \textbf{$b\to a$}, we get back to a FLRW manifold. So 
we have a smooth transition to a homogeneous
and isotropic universe. In the following we will only consider the
vacuum case. In this study we perform the same analysis as in
\cite{DeFelice:2013awa}. In particular, in GR one would expect that
only one odd mode is propagating, namely one of the two
polarizations of gravitational waves.

\section{The background and the FLRW limit\label{back_FLRWlimit}}

In this section, we show the existence of a smooth isotropic limit
for Bianchi-I background solutions in ECG gravity. We need such a
solution in order to understand the behavior of the extra modes,
present in the theory, in the smooth isotropic FLRW limit of a Bianchi
type I metric. 
If there were no such stable isotropic limit, this would not be possible.

In order to achieve this goal, we try then to solve the equations of
background motion iteratively in the isotropic limit. In particular, we have two
differential equations for the two background variables $a$ and $b$,
and when $b\to a$ we have a background which reduces to FLRW, whose
solution describes the de Sitter expansion. Therefore we can try to find an
iterative solution of the following kind
\begin{eqnarray}
a & = & a_{0}+a_{1}+\dots\,,\\
b & = & a_{0}+b_{1}+\dots\,,
\end{eqnarray}
where $a_{1},b_{1}\ll a_{0}$, and 
\begin{equation}
a_{0}\propto e^{H_{0}t}\,,
\end{equation}
where $H_0$ is given as the solution of the first Friedmann equation,
namely Eq.\ (\ref{eq:1stFRIED}).  Since we know that $a_{0}$ satisfies
the equations of motion for the FLRW background, we can linearize the
background equations for $a_{1}$ and $b_{1}$ and find their
dynamics. In principle, one can continue further to the next order with the
condition that $a_{2},b_{2}\ll a_{1},b_{1}\ll a_{0}$, etc. If it is
possible to build iteratively such a solution, then indeed we can construct
an anisotropic Bianchi-I background which approximates an isotropic FLRW one.

Actually, we have three equations of motion in Bianchi-I corresponding to the 
lapse equation and the two equations of motion for the fields $a$ and $b$. 
However, because of the Bianchi identities,
only two of them are independent. At the lowest order, as we already know 
by construction of such a solution, one finds that the $a$-equation of motion implies
\begin{equation}
a_{0}\ddot{a}_{0}-\dot{a}_{0}^{2}=0\,,
\end{equation}
which is automatically satisfied on de Sitter. This is the analogue of
the GR equation $\dot{H}=-4\pi G(\rho+p)=0$, in the presence of only a
cosmological constant. Among all family of de Sitter solutions,
we impose here the condition
\begin{equation}
\Mpl^{4}+48\beta\,H_{0}^{4}>0\,,\label{eq:no-ghost-FLRW}
\end{equation}
otherwise we would have that the tensor modes on FLRW either become
(massless) ghosts (for $\beta<-\frac{1}{48}\Mpl^{4}/H_{0}^{4}$) or
they get strongly coupled (for
$\beta=-\frac{1}{48}\Mpl^{4}/H_{0}^{4}$).  In the following we will
not consider this nonphysical situation.  Furthermore, for obvious
reasons, we will consider the case $\beta\neq0$.  At the lowest order
the $b$-equation of motion (see the appendix
\ref{sec:Background-equations-of} for more details) does not add any
new information.

Now that we have set the zeroth order solution, we can proceed to
find the deviations from exact de Sitter by studying the next variables
$a_{1}$ and $b_{1}$.

At the first order in the variables $a_{1}$, and $b_{1}$, we get
from the $\Lambda$-equation 
\begin{equation}
\dot{a}_{1}+2\dot{b}_{1}=H_{0}\,(a_{1}+2b_{1})\,.
\end{equation}
The solution for the previous equation can be written as 
\begin{equation}
b_{1}=-\frac{1}{2}\,a_{1}+\mathcal{C}_{1}\,e^{H_{0}t}\,,
\end{equation}
where the last term containing $\mathcal{C}_{1}$ can be included into a
renormalization of the background lowest order solution for which
$a_{0}\propto e^{H_{0}t}$. Therefore we can set
\begin{equation}
b_{1}=-\frac{1}{2}\,a_{1}\,.
\end{equation}
The $b$-equation instead gives 
\begin{equation}
\ddot{a}_{1}+H_{0}\dot{a}_{1}-2H_{0}^{2}\,a_{1}=0\,,
\end{equation}
which is solved by 
\begin{equation}
a_{1}=-2\mathcal{C}_{1}\,e^{-2H_{0}t}+\mathcal{C}_{2}\,e^{H_{0}t}\,.
\end{equation}
Once more the second solution can be reabsorbed into the lowest order
term, so that we are left with 
\begin{eqnarray}
a_{1} & = & -2\mathcal{C}_{1}\,e^{-2H_{0}t}=-\frac{2C_{1}}{a_{0}^{2}}\,,\\
b_{1} & = & \mathcal{C}_{1}\,e^{-2H_{0}t}=\frac{C_{1}}{a_{0}^{2}}\,.
\end{eqnarray}
Indeed we find that, for a large range of values of $C_{1}$,
$a_{1},b_{1}\ll a_{0}$ when $a_{0}>1$ as expected during inflation. 
The other remaining equation of motion is automatically
satisfied at the same order, because of Bianchi identities.

At the next order one finds from the $\Lambda$-equation 
\[
[(2\dot{b}_{2}+\dot{a}_{2})-H_{0}(2b_{2}+a_{2})]\,a_{0}^{5}\left(\frac{\Mpl^{4}}{48}+\beta H_{0}^{4}\right)+\frac{387C_{1}^{2}H_{0}}{2}\left(\frac{\Mpl^{4}}{2064}+\beta H_{0}^{4}\right)=0\,,
\]
which is solved by 
\begin{equation}
a_{2}=-2b_{2}+\frac{3C_{1}^{2}}{4\,a_{0}^{5}}\left(\frac{\Mpl^{4}+2064\beta H_{0}^{4}}{\Mpl^{4}+48\beta H_{0}^{4}}\right),
\end{equation}
where we have once more discarded a term proportional to $e^{H_{0}t}$.
The $b$-equation of motion gives 
\begin{equation}
b_{2}=-\frac{C_{1}^{2}}{4\,a_{0}^{5}}\left(\frac{\Mpl^{4}+2064\beta H_{0}^{4}}{\Mpl^{4}+48\beta H_{0}^{4}}\right),
\end{equation}
where we have discarded two terms proportional to $e^{H_{0}t}$ and
$e^{-2H_{0}t}$ respectively as renormalizations of the previous $a_{0}$
and $a_{1}$ solutions. Therefore we find 
\[
a_{2}=-5b_{2}.
\]
In this way one we have found a solution for which
$(a_{2},b_{2})\propto a_0^{-5}\ll (a_{1},b_{1})\propto a_0^{-2}\ll
a_{0}\propto e^{H_0 t}$, and one can continue building up such a
solution order by order in inverse powers of $a_0$. Finally, we have
shown the existence of a solution (at least up to second order) for
which $a_{0}$ grows exponentially in time, and 
$\lim_{a_{0}\to\infty}b(t)=a(t)=a_{0}(t)$.  Up to the second order
(but we may continue further on), we find that this approximate
solution can be written as
\begin{eqnarray}
a & = & a_{0}\left[1-\frac{2C_{1}}{a_{0}^{3}}+\frac{5C_{1}^{2}}{4\,a_{0}^{6}}\left(\frac{\Mpl^{4}+2064\beta H_{0}^{4}}{\Mpl^{4}+48\beta H_{0}^{4}}\right)+\mathcal{O}(C_{1}^{3}/a_{0}^{9})\right],\\
b & = & a_{0}\left[1+\frac{C_{1}}{a_{0}^{3}}-\frac{C_{1}^{2}}{4\,a_{0}^{6}}\left(\frac{\Mpl^{4}+2064\beta H_{0}^{4}}{\Mpl^{4}+48\beta H_{0}^{4}}\right)+\mathcal{O}(C_{1}^{3}/a_{0}^{9})\right].
\end{eqnarray}

\section{Degrees of freedom for odd modes \label{dof}}

We expand the Lagrangian density at the second order in perturbations for
a general Bianchi-I background for the theory under consideration.
Then we will consider the limit for $b\to a$. Since the background
is homogeneous, we can expand the perturbation variables in Fourier
modes also for the $x$ coordinate, with a basis which satisfies $\partial_{x}^{2}\tilde{Y}(x)=-k^{2}\tilde{Y}(x)$.
Here we will consider only the constraints coming from the local behavior
of linear perturbations. For this aim, for simplicity, but without
lack of generality, we can focus on one single Fourier mode with the wave
vector $\vec{K}=(k,\vec{q})$.

We have three background equations of motion (but only two of them
are independent). On imposing the background equations of motion on
the second order Lagrangian density $\mathcal{L}$, we notice that the following terms (which
are new compared to GR) come out, namely

\begin{equation}
\mathcal{L}=\frac{6\beta\dot{b}k^{2}q^{2}}{\Mpl^{2}b^{2}}\,(a\dot{b}-b\dot{a})\,\ddot{W}\left(\ddot{W}-\frac{2}{a}\,\dot{V}\right)+\frac{6\beta q^{2}\dot{b}(k^{2}b^{2}-2q^{2}a^{2})}{\Mpl^{2}b^{4}a^{2}}\,(a\dot{b}-b\dot{a})\,\dot{V}^{2}+\dots,
\end{equation}
These terms are absent in GR, because \textbf{$\beta=0$} identically.
Then in GR, $V$ would represent a Lagrange multiplier which
can be integrated out leaving a reduced action for a single propagating
mode, $W$, which would have a standard kinetic term. It is interesting
to notice that in the exact isotropic limit i.e.\ $b\propto a,$ the
above terms in the action vanish, hence the mode $V$ becomes
a Lagrangian multiplier on FLRW.

However, for this new theory, i.e.\ $\beta\neq0$, the presence of these 
two terms in the non-isotropic case suggests that new degrees of
freedom will arise in general. The highest-derivative terms responsible
for the presence of the new modes, on the other hand, tend to vanish
in the isotropic limit, so that we need to understand what happens
to such degrees of freedom in this limit.

We can rewrite the previous terms in the following equivalent way 
\begin{equation}
\mathcal{L}=\frac{6\beta\dot{b}k^{2}q^{2}}{\Mpl^{2}b^{2}}\,(a\dot{b}-b\dot{a})\left(\ddot{W}-\frac{1}{a}\,\dot{V}\right)^{2}-\frac{12\beta q^{4}\dot{b}}{\Mpl^{2}b^{4}}\,(a\dot{b}-b\dot{a})\,\dot{V}^{2}+\dots,
\end{equation}
and the terms above can be rewritten as 
\begin{equation}
\mathcal{L}'=\frac{6\beta\dot{b}k^{2}q^{2}}{\Mpl^{2}b^{2}}\,(a\dot{b}-b\dot{a})\left[2\zeta\left(\ddot{W}-\frac{1}{a}\,\dot{V}\right)-\zeta^{2}\right]-\frac{12\beta q^{4}\dot{b}}{\Mpl^{2}b^{4}}\,(a\dot{b}-b\dot{a})\,\dot{V}^{2}+\dots,
\end{equation}
where we have included a new Lagrange multiplier $\zeta$, whose equation
of motion (algebraic in $\zeta$ itself) is 
\begin{equation}
\zeta=\ddot{W}-\frac{1}{a}\,\dot{V}\,.
\end{equation}
In fact, on replacing it inside the new Lagrangian density $\mathcal{L}'$, we get the original
Lagrangian density $\mathcal{L}$. In other words, we have two equivalent Lagrangian densities which lead
to the same classical equations of motion.

After integrating by parts the term involving $\ddot{W}$, we get
a term of the form $\dot{\zeta}\dot{W}$, so that in the new obtained
Lagrangian density $\mathcal{L}'$ we can represent now the kinetic term for the fields $\psi_{i}=(\zeta,W,V)$
as 
\begin{equation}
\mathcal{L}'=K_{ij}\dot{\psi}_{i}\dot{\psi}_{j}+\dots,
\end{equation}
where $K_{ij}=K_{ji}$, $K_{11}=0=K_{13}$, but $K_{12}\neq0$. In
the exact isotropic limit, the terms $K_{12}$, $K_{23}$, $K_{33}$
all vanish, making the field $V=\psi_{3}$ a Lagrange multiplier.
We now try to diagonalize the matrix $K_{ij}$ by the following
field redefinition 
\begin{eqnarray}
\zeta & = & F_{1}\,,\\
W & = & \Gamma_{1}\,F_{1}+F_{2}\,,\\
V & = & \Gamma_{2}\,F_{1}+\Gamma_{3}\,F_{2}+F_{3}\,.
\end{eqnarray}
This transformation is in general well defined (as shown in the appendix
\ref{sec:Regular-field-redefinition}), even in the isotropic limit,
for example 
\begin{equation}
\Gamma_{3}=\frac{k^{2}b}{q^{2}a^{2}}\,(a\dot{b}-\dot{a}b)\,,
\end{equation}
and its determinant is equal to unity. After this field redefinition,
it is possible to write down the new diagonal kinetic matrix $A_{ij}$
as follows 
\begin{equation}
\mathcal{L}'=A_{ij}\,\dot{F}_{i}\dot{F}_{j}+\dots,
\end{equation}
with three different diagonal elements. Investigation of the positivity of
these diagonal elements is sufficient for understanding whether the theory
has ghosts or not. In fact, we find 
\begin{eqnarray}
A_{11}=g_{1} & = & -\frac{K_{33}K_{12}^{2}}{K_{22}K_{33}-K_{23}^{2}}\,,\\
A_{22}=g_{2} & = & \frac{K_{22}K_{33}-K_{23}^{2}}{K_{33}}\,,\\
A_{33}=g_{3} & = & K_{33}=\frac{12\,(\dot{a}b-a\dot{b})\,\beta\dot{b}\,q^{4}}{b^{4}\Mpl^{2}}\,,
\end{eqnarray}
from which it is clear that $g_{1}g_{2}=-K_{12}^{2}<0$. Therefore,
no matter what the evolution is, there will be always at least one ghost
mode in the odd sector. If also $K_{33}<0$, then two ghosts will
be present. Furthermore, both $g_{1}$ and $g_{3}$ tend to vanish
in the isotropic limit, whereas $g_{2}$ does not. This was expected,
as on FLRW we should apparently get only one odd propagating
mode.

For the solution approaching FLRW which was found in the previous
section, we find 
\begin{eqnarray}
g_{1} & = & -\frac{11664\,k^{2}q^{2}H_{0}^{4}\,\beta^{2}C_{1}^{2}}{a_{0}^{5}\Mpl^{2}\,(\Mpl^{4}+48\,\beta\,H_{0}^{4})}+\mathcal{O}(C_{1}^{3}/a_{0}^{8})\,,\\
g_{2} & = & \frac{k^{2}q^{2}a_{0}\,(\Mpl^{4}+48\,\beta\,H_{0}^{4})}{4\Mpl^{2}}+\mathcal{O}(C_{1}/a_{0}^{2})\,,\\
g_{3} & = & \frac{108q^{4}H_{0}^{2}\,\beta\,C_{1}}{a_{0}^{4}\Mpl^{2}}+\mathcal{O}(C_{1}^{2}/a_{0}^{7})\,.
\end{eqnarray}
By investigating these expressions we can see that, for all allowed
values of $\beta$, $g_{1}<0$, whereas $g_{2}>0$, i.e.\ the field
$F_{2}$ is always well behaved. In fact, since $F_{2}$ is the only
odd mode which seems to exist on exact FLRW, it should represent a
tensor mode. If $C_{1}\beta<0$, then $F_{3}$ represents a (second)
ghost. We will distinguish the two cases which depend on the sign
of $\beta C_{1}$. If $\beta C_{1}>0$, then $F_{2}$ and $F_{3}$
are never ghost degrees of freedom in the isotropic limit. If one
takes the exact FLRW limit, one re-obtains the standard equation of
motion for one polarization (the cross one) for the de Sitter cosmological
tensor modes, as also shown in the Appendix \ref{sec:Regular-field-redefinition}.

\section{Mass dispersion relations\label{mdr}}

In the following we want to study behavior of the perturbations
variables, in particular we want to see what happens to them in the
isotropic limit.

\subsection{One ghost case}

In this case we choose $\beta C_{1}>0$, that is $g_{3}>0$, whereas
$A_{11}<0$, so that only one ghost exists. In order to study canonically 
normalized fields on FLRW, we can make another field
redefinition by imposing 
\begin{eqnarray}
F_{1} & = & \frac{a_{0}^{3/2}}{\sqrt{-2A_{11}}}\,Z_{1}\,,\\
F_{2} & = & \frac{a_{0}^{3/2}}{\sqrt{2A_{22}}}\,Z_{2}\,,\\
F_{3} & = & \frac{a_{0}^{3/2}}{\sqrt{2A_{33}}}\,Z_{3}\,,
\end{eqnarray}
so that the Lagrangian density for the perturbations can be rewritten as 
\begin{equation}
\mathcal{L}_{{\rm odd}}=\frac{a_{0}^{3}}{2}\left[-\dot{Z}_{1}^{2}+\dot{Z}_{2}^{2}+\dot{Z}_{3}^{2}+B_{ij}\,(\dot{Z}_{i}Z_{j}-Z_{i}\dot{Z}_{j})-\mu_{ij}\,Z_{i}Z_{j}\right].\label{eq:Lag_one_ghost}
\end{equation}
Now it is clear that in the isotropic limit, when the both $A_{11}$
and $A_{33}$ tend to vanish, then in general other terms in the
Lagrangian density might become larger and larger, in particular the mass term
of such modes. If this mass will be positive, then the mode would
become very massive, but if such a mass is negative this mode (if not
a ghost) would become extremely unstable. In such a situation, as we
shall see later on, the theory develops a short-time instability,
i.e.\ an instability which cannot be neglected in a Hubble time.

In order to see what happens to the stability of perturbations
by evaluating their mass-dispersion relation, we should take equations 
for them which look as follows
\begin{eqnarray}
  \ddot{Z}_1&+&\mu_{1j}Z_j+2B_{1j}\,\dot{Z}_j+\dots=0\,,\\
  \ddot{Z}_2&-&\mu_{2j}Z_j-2B_{2j}\,\dot{Z}_j+\dots=0\,,\\
  \ddot{Z}_3&-&\mu_{3j}Z_j-2B_{3j}\,\dot{Z}_j+\dots=0
\end{eqnarray}  
and consider  the isotropic limit for the anti-symmetric
matrix $B_{ij}$ and for the symmetric matrix $\mu_{ij}$. One can see
that at the lowest order in isotropy, one finds
\begin{eqnarray}
B_{12} & = & -\frac{H_{0}}{2}+\mathcal{O}(C_{1}/a_{0}^{3})\,,\\
B_{13} & = & \frac{3kH_{0}\sqrt{3\beta C_{1}}\,(H_{0}^{2}+2q^{2}/a_{0}^{2})}{a_{0}^{3/2}q\sqrt{\Mpl^{4}+48\,\beta\,H_{0}^{4}}}+\mathcal{O}[(\beta C_{1}/a_{0}^{3})^{3/2}]\,,\\
B_{23} & = & -\frac{a_{0}^{3/2}k\sqrt{3(\Mpl^{4}+48\,\beta\,H_{0}^{4})}}{72\,\sqrt{\beta C_{1}}H_{0}q}+\mathcal{O}[(\beta C_{1}/a_{0}^{3})^{1/2}]\,.
\end{eqnarray}
Then the eigenvalues of the matrix $\mu_{ij}$ determine the mass
eigenvalues of the modes. We find that in the isotropic limit the
elements of the matrix $\mu_{ij}$ reduce to: 
\begin{eqnarray}
\mu_{11} & = & -\frac{(\Mpl^{4}+48\,\beta\,H_{0}^{4})\,a_{0}^{3}}{216\,H_{0}^{2}\,\beta\,C_{1}}-\frac{(\Mpl^{4}+1248\,\beta\,H_{0}^{4})+72\,H_{0}^{2}\beta\,(k^{2}+q^{2})/a_{0}^{2}}{72\,H_{0}^{2}\beta}\,,\\
\mu_{22} & = & \frac{q^{2}}{a_{0}^{2}}+\mathcal{O}(C_{1}/a_{0}^{3})\,,\\
\mu_{33} & = & -\frac{(k^{2}+q^{2})(\Mpl^{4}+48\,\beta\,H_{0}^{4})\,a_{0}^{3}}{432\,C_{1}q^{2}H_{0}^{2}\beta}-\frac{[(1128\,k^{2}+348\,q^{2})H_{0}^{4}\beta+\Mpl^{4}(3\,k^{2}+q^{2})]a_{0}^{2}+72\,(k^{2}+q^{2})^{2}H_{0}^{2}\beta}{144\,H_{0}^{2}a_{0}^{2}\beta\,q^{2}}\,,\\
\mu_{12} & = & -\frac{q^{2}}{a_{0}^{2}}+\frac{5}{2}\,H_{0}^{2}+\mathcal{O}(C_{1}/a_{0}^{3})\,,\\
\mu_{13} & = & \frac{a_{0}^{3/2}k\,\sqrt{3(\Mpl^{4}+48\,\beta\,H_{0}^{4})}}{36\sqrt{\beta C_{1}}q}+\mathcal{O}[(\beta C_{1}/a_{0}^{3})^{1/2}]\,,\\
\mu_{23} & = & -\frac{a_{0}^{3/2}k\sqrt{3(\Mpl^{4}+48\,\beta\,H_{0}^{4})}}{16\sqrt{\beta C_{1}}q}+\mathcal{O}[(\beta C_{1}/a_{0}^{3})^{1/2}]\,.
\end{eqnarray}
In other words, we can see that the mass matrix can be approximated
as 
\begin{equation}
\mu_{ij}=\frac{\Mpl^{4}+48\,\beta\,H_{0}^{4}}{432H_{0}^{2}}\,\frac{a_{0}^{3}}{\beta C_{1}}\left(\begin{array}{ccc}
-2 & \mathcal{\mathcal{O}}(C_{1}/a_{0}^{3}) & \mathcal{\mathcal{O}}(\sqrt{\beta C_{1}/a_{0}^{3}})\\
\mathcal{\mathcal{O}}(C_{1}/a_{0}^{3}) & \mathcal{\mathcal{O}}(C_{1}/a_{0}^{3}) & \mathcal{\mathcal{O}}(\sqrt{\beta C_{1}/a_{0}^{3}})\\
\mathcal{\mathcal{O}}(\sqrt{\beta C_{1}/a_{0}^{3}}) & \mathcal{\mathcal{O}}(\sqrt{\beta C_{1}/a_{0}^{3}}) & -\frac{k^{2}+q^{2}}{q^{2}}
\end{array}\right),
\end{equation}
so that to the lowest order, the modes $Z_{1}$ and $Z_{3}$ have
negative self-coupling terms. For the ghost mode, this is actually a
good point, because it would make it stable\footnote{In fact, for a
  stable harmonic oscillator we have $L=\dot{x}^{2}-\omega^{2}x^{2}$,
  whereas for a stable ghost we should have
  $L_{g}=-\dot{x}^{2}+\omega^{2}x^{2}=-L$, because both $L$ and
  $L_{g}$ lead to the same equations of motion.}. However the $Z_{3}$
mode, which is not a ghost, tends to be strongly unstable in the
isotropic limit. Notice that $(k^{2}+q^{2})/q^{2}>1$, so that this
problem takes place at any scale (and gets worse when $q/k\to0$). As
already noticed above, the reason why e.g.\ the term $\mu_{33}$
becomes larger and larger in the isotropic limit is due to the fact
that the coefficient of $\dot{F}_{33}^2$ tends to vanish in the
same limit.

This instability is purely classical, so that we do not need to invoke
any quantum particle production. That is due to the fact that $Z_{3}$
becomes a tachyon, its mass growing exponentially but towards more and
more negative values. Thus, we expect to have an exponentially growing
instability when we solve the equations of motion.

\subsection{Two ghosts case}

Along the same lines as in the previous section, in this case we consider
$C_{2}\equiv-C_{1}$, together with $\beta C_{2}>0$.  In this case, we
find that in the isotropic limit both $g_{1}$ and $g_{3}$ become
negative, so that there are actually two ghost degrees of freedom. We can make
a further field redefinition
\begin{eqnarray}
F_{1} & = & \frac{a_{0}^{3/2}}{\sqrt{-2A_{11}}}\,Z_{1}\,,\\
F_{2} & = & \frac{a_{0}^{3/2}}{\sqrt{2A_{22}}}\,Z_{2}\,,\\
F_{3} & = & \frac{a_{0}^{3/2}}{\sqrt{-2A_{33}}}\,Z_{3}\,,
\end{eqnarray}
which is convenient in the isotropic limit, so that the Lagrangian density
for the perturbations can be rewritten as 
\begin{equation}
\mathcal{L}_{{\rm odd}}=\frac{a_{0}^{3}}{2}\left[-\dot{Z}_{1}^{2}+\dot{Z}_{2}^{2}-\dot{Z}_{3}^{2}+B_{ij}(\dot{Z}_{i}Z_{j}-Z_{i}\dot{Z}_{j})-\mu_{ij}\,Z_{i}Z_{j}\right].\label{eq:Lag_two_ghosts}
\end{equation}
Then we consider the isotropic limit for the anti-symmetric matrix
$B_{ij}$ and for the symmetric matrix $\mu_{ij}$, which then become
functions of $a_{0}$, the wave numbers $q$ and $k$, the parameters
of the action, and finally of $C_{2}$. One can see that at the lowest
order in isotropy, one finds 
\begin{eqnarray}
B_{12} & = & \frac{H_{0}}{2}+\mathcal{O}(C_{2}/a_{0}^{3})\,,\\
B_{13} & = & \frac{3kH_{0}\sqrt{3\beta C_{2}}\,(H_{0}^{2}+2q^{2}/a_{0}^{2})}{a_{0}^{3/2}q\sqrt{\Mpl^{4}+48\,\beta\,H_{0}^{4}}}+\mathcal{O}[(\beta C_{2}/a_{0}^{3})^{3/2}])\,,\\
B_{23} & = & -\frac{a_{0}^{3/2}k\sqrt{3(\Mpl^{4}+48\,\beta\,H_{0}^{4})}}{72\,\sqrt{\beta C_{2}}H_{0}q}+\mathcal{O}[(\beta C_{2}/a_{0}^{3})^{1/2}]\,.
\end{eqnarray}
Then the eigenvalues of the matrix $\mu_{ij}$ determine the mass
eigenvalues of the modes. We find that in the isotropic limit the
element of matrix $\mu_{ij}$ reduce to: 
\begin{eqnarray}
\mu_{11} & = & \frac{(\Mpl^{4}+48\,\beta\,H_{0}^{4})\,a_{0}^{3}}{216\,H_{0}^{2}\beta\,C_{2}}-\frac{\left(\Mpl^{4}+1248\,\beta\,H_{0}^{4}\right)+72\,H_{0}^{2}\beta\,\left(k^{2}+q^{2}\right)/a_{0}^{2}}{72\,H_{0}^{2}\beta}\,,\\
\mu_{22} & = & \frac{q^{2}}{a_{0}^{2}}+\mathcal{O}(C_{2}/a_{0}^{3})\,,\\
\mu_{33} & = & -\frac{(k^{2}+q^{2})(\Mpl^{4}+48\,\beta\,H_{0}^{4})a_{0}^{3}}{432\,C_{2}q^{2}H_{0}^{2}\beta}+\frac{[(1128\,k^{2}+348\,q^{2})H_{0}^{4}\beta+\Mpl^{4}(3\,k^{2}+q^{2})]a_{0}^{2}+72\,(k^{2}+q^{2})^{2}H_{0}^{2}\beta}{144\,H_{0}^{2}a_{0}^{2}\beta\,q^{2}}\,,\\
\mu_{12} & = & \frac{q^{2}}{a_{0}^{2}}-\frac{5}{2}\,H_{0}^{2}+\mathcal{O}(C_{2}/a_{0}^{3})\,,\\
\mu_{13} & = & -\frac{a_{0}^{3/2}k\sqrt{3(\Mpl^{4}+48\,\beta\,H_{0}^{4})}}{36\sqrt{\beta C_{{2}}}q}+\mathcal{O}[(\beta C_{2}/a_{0}^{3})^{1/2}]\,,\\
\mu_{23} & = & -\frac{a_{0}^{3/2}k\sqrt{3(\Mpl^{4}+48\,\beta\,H_{0}^{4})}}{16\sqrt{\beta C_{2}}q}+\mathcal{O}[(\beta C_{2}/a_{0}^{3})^{1/2}]\,.
\end{eqnarray}
In this case, we can see that the mass matrix can be approximated
as 
\begin{equation}
\mu_{ij}=\frac{\Mpl^{4}+48\,\beta\,H_{0}^{4}}{432H_{0}^{2}}\,\frac{a_{0}^{3}}{\beta C_{2}}\left(\begin{array}{ccc}
2 & \mathcal{\mathcal{O}}(C_{2}/a_{0}^{3}) & \mathcal{\mathcal{O}}(\sqrt{\beta C_{2}/a_{0}^{3}})\\
\mathcal{\mathcal{O}}(C_{2}/a_{0}^{3}) & \mathcal{\mathcal{O}}(C_{2}/a_{0}^{3}) & \mathcal{\mathcal{O}}(\sqrt{\beta C_{2}/a_{0}^{3}})\\
\mathcal{\mathcal{O}}(\sqrt{\beta C_{2}/a_{0}^{3}}) & \mathcal{\mathcal{O}}(\sqrt{\beta C_{2}/a_{0}^{3}}) & -\frac{k^{2}+q^{2}}{q^{2}}
\end{array}\right),
\end{equation}
so that to the lowest order the modes $Z_{1}$ and $Z_{3}$ have 
self-coupling terms of opposite signs. For the ghost mode $Z_{1}$, a 
positive squared-mass diagonal element $\mu_{11}$  corresponds to a tachyonic instability. In this case a negative
mass for $Z_{3}$, which is now also a ghost, would instead make it
stable. In any case, the whole system, because of the tachyonic mass
for the ghost, tends to be unstable. As in the one ghost case, this instability 
is purely classical, so that we do not need to invoke any
quantum particle production, which could of course contribute to produce
an additional instability. However, even from a pure classical level,
the background will be unstable. Thus, we expect to have an exponentially 
growing instability in this case, too,  when we solve the equations of
motion.

\section{Numerical integration\label{num}}

We want to show numerically that the instability studied analytically in the previous section
develops well before even one single e-folding in general. This
implies we should expect the de Sitter solution to be unstable before
the effective field theory approach breaks down. In order to check the
characteristic time of such an instability, in the following, we will solve
numerically the full equations of motion for the two Lagrangian densities
described in the Eqs.\ (\ref{eq:Lag_one_ghost}) and
(\ref{eq:Lag_two_ghosts}), assuming
\begin{eqnarray}
a & \approx & a_{0}-\frac{2C_{1}}{a_{0}^{2}}\,,\\
b & \approx & a_{0}+\frac{C_{1}}{a_{0}^{2}}\,.
\end{eqnarray}
On replacing $q=\bar{q}H_{0}$, $k=\bar{k}H_{0}$, and
$H_{0}=\alpha\Mpl$, we integrate the equations of motion with respect
to the number of e-folds $N\equiv\ln a_{0}/a_{0,{\rm ini}}$ variable by
adding the extra equation of motion $a'_{0}=a_{0}$, so that the system
of ODEs becomes autonomous, i.e.\ explicitly independent of $N$. We
have considered typical values for both parameters and initial
conditions ($Z_{i,{\rm ini}}=10^{-6}$, $\dot{Z}_{i,{\rm ini}}=0$,
$a_{0,{\rm ini}}=1$).  The results are shown in Figure 1 and they
confirm the analytical prediction for the existence of a classical
instability. Here, we have chosen these initial conditions which looks
sensible. Indeed we want to start from a universe which is close to a
FLRW one, and see what happens next to the perturbations variables. 
Even when we change the initial conditions, we get a similar unstable
behavior. This provides additional support for a generic exponential 
growth of such an instability, making the FLRW behavior nonviable. 
We have checked that extending the expansion of solutions for $a$ 
and $b$ to the order $\mathcal{O}(C_{1}^{2}/a_{0}^{6})$ does not 
change numerical results qualitatively.

\begin{figure}
\subfloat[Classical instability in the single ghost case, that is $\beta C_{1}>0$.]{\includegraphics[width=9cm]{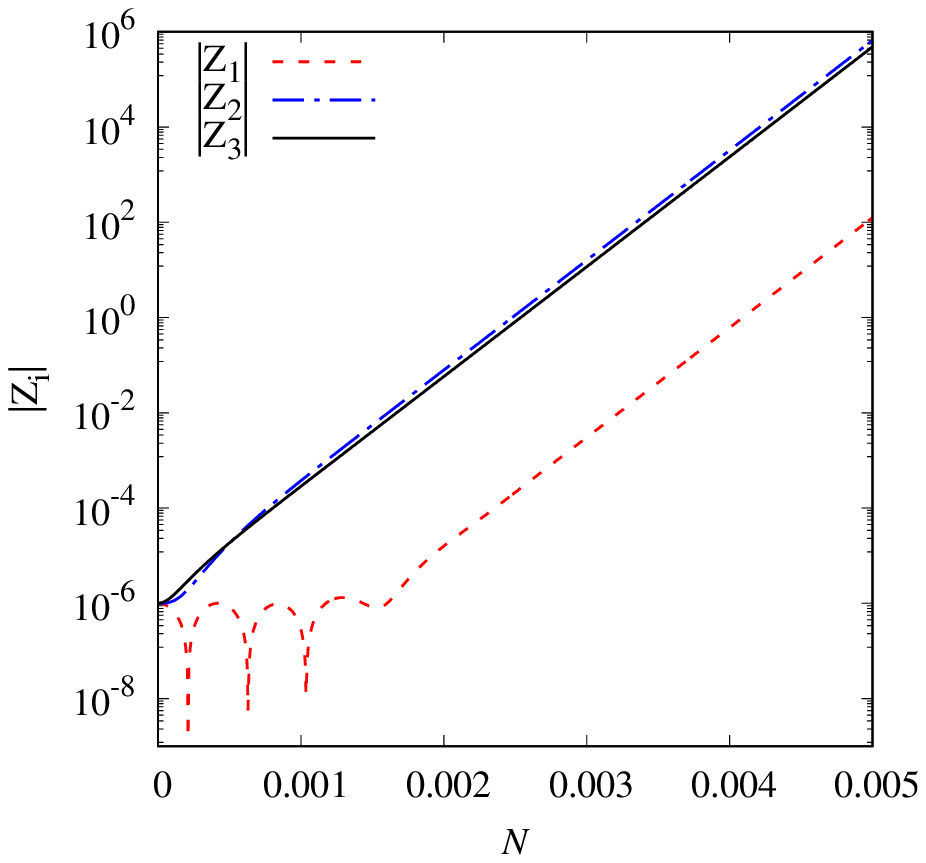}

}\subfloat[Classical instability in the double ghost case, that is $\beta C_{1}<0$.]{\includegraphics[width=9cm]{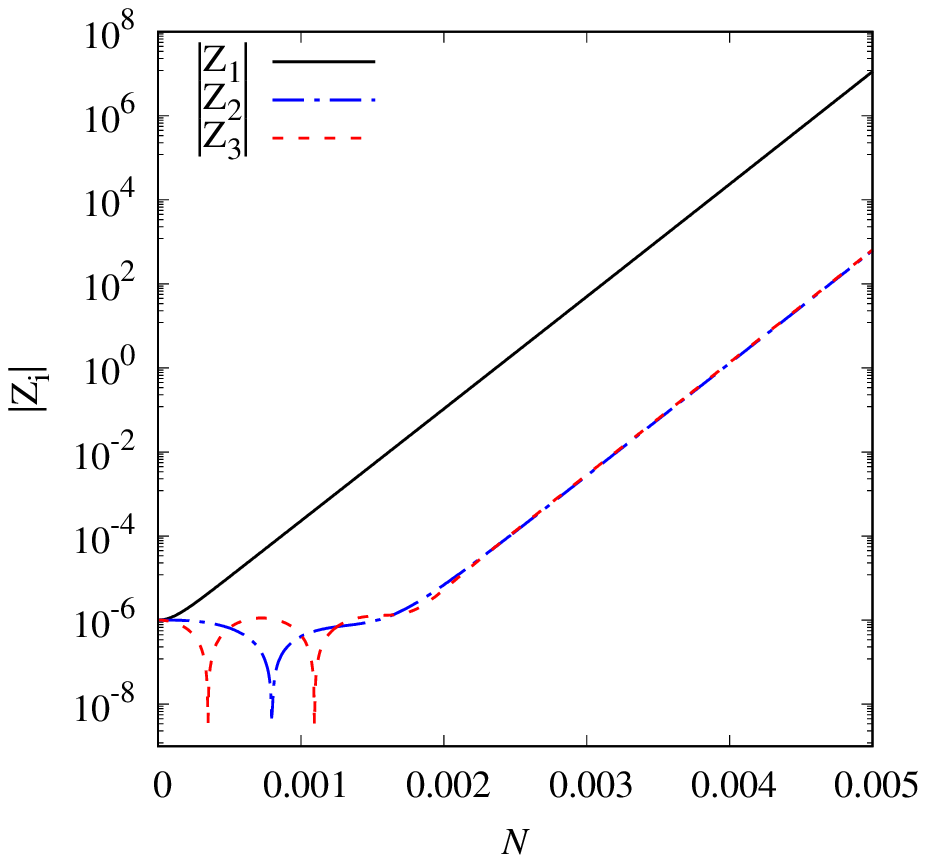}

}\caption{Classical instability present in the theory in the isotropic limit,
when $b\to a$. In the left panel, we have set $H_{0}/\Mpl=10^{-2}$,
$q=k=10H_{0}$, $\beta=0.1=C_{1}$. We can see that the non-ghost
mode $Z_{3}$ is exponentially unstable and its growth makes 
the other modes grow exponentially. In the right panel, in the case
with two ghosts, we have set $\frac{H_{0}}{\Mpl}=10^{-2}$, $\frac{q}{3}=10H_{0}$,
$k=10H_{0}$, $\beta=0.1=-C_{1}$. Here we can see that the ghost
mode $Z_{1}$ is exponentially unstable and its growth makes 
the other modes grow exponentially, too. \hfill{}\hfill{}}
\end{figure}

\section{Conclusions\label{conclusion}}

After introducing the ECG theory \cite{Bueno:2016xff,Arciniega:2018tnn}
a generalization of this theory has been also proposed and its cosmology
has been studied at the background level \cite{Erices:2019mkd}. In
\cite{Erices:2019mkd}, it is proposed that this theory can explain
both early universe inflationary era and the late-time acceleration
of the universe. It was shown that the theory does not possess any
ghost modes on a FLRW background. However, as we show here,
this is not enough to say a theory is viable or not. One has to investigate
its linear perturbations -- and not only on a FLRW background
-- to show that the theory does not possess any instabilities or
strong coupling behavior. 

Since ECG is a higher order theory in the curvature tensor, as predicted
by the Lovelock theorem, it should contain extra degrees of freedom.
On the other hand, in this theory only two degrees of freedom propagate 
on a FLRW manifold in the absence of matter, as in GR. In order to look 
at the nature of other degrees of freedom which present in ECG and 
whose existence is predicted by the Lovelock theorem, we have studied 
linear perturbations on a homogeneous anisotropic Bianchi type I 
spacetime in this gravity theory. Such consideration introduces an 
anisotropy (with the infinite length scale) already at the background 
level compared to the isotropic FLRW metric. We showed the existence 
of a vacuum solution for Bianchi-I background which smoothly approaches 
the de Sitter metric in the isotropic limit, and for which anisotropy 
decreases with time as $a_{0}^{-3}$, where $a_{0}\propto e^{H_{0}t}$.
Therefore, we can study the FLRW limit of a Bianchi-I manifold.
As it was known in literature, we re-obtained that on a FLRW background,
the ECG theory should satisfy the no-ghost condition $\Mpl^{4}+48\beta H_{0}^{4}>0$
for tensor modes of perturbations. Together with the background Friedman 
equation, it is interesting to see that this theory always requires 
$\Lambda>2H_{0}^2>0$ that indicates the absence of stable de Sitter 
solutions without a bare positive cosmological constant of the order of 
$H_{0}^2$.

However, studying inhomogeneous perturbations on a Bianchi-I background
in the regime of small anisotropy, we have expanded the Lagrangian density 
up to the second order in odd-parity perturbations and have found the 
existence of new degrees of freedom (compared to GR), confirming the 
predictions of the Lovelock theorem. The kinetic term of these new 
degrees vanishes in the exact isotropic limit. Thus, consideration of
linear perturbations on a Bianchi-I manifold, even in its isotropic limit, 
makes possible to go beyond the linear order with respect to a FLRW 
background and to obtain non-perturbative results. We find that in 
total, as for odd modes, three degrees of freedoms are present. Then 
we diagonalize the kinetic matrix and redefine fields into canonically 
normalized ones in order to study the speed of their propagation and 
mass dispersion relations.

We find that for any parameter of ECG theory, there always exists at least
one ghost which propagates in the background. Furthermore, we also find
that one of the modes always acquire a tachyonic instability with
time-scale much shorter than the Hubble time. We checked numerically 
that the instability grows well before one single e-fold. The instability
present here is pure classical, thus the quantum particle production
needs not be studied. 

This study shows that ECG theory has a classical instability in the isotropic 
de Sitter limit, which is present at all scales and is developed even before 
one e-fold. Thus, without curing this problem, this theory cannot be 
considered as a viable theory of gravity and cannot be used to construct 
internally consistent isotropic cosmological models. Note finally that this 
phenomenon is very similar to that arising in $f(R,G)$ theory of gravity, 
where $G$ is the Gauss-Bonnet invariant, and making it not viable, too, 
apart from some specific exceptional cases~\cite{DeFelice:2010hg}.

\appendix

\section{Background equations of motion\label{sec:Background-equations-of}}

We write in the following the equations of motion for the background.
We have three equations of motion, which can be written as follows
\begin{eqnarray}
E_{1} & \equiv & -\frac{\Lambda}{6}+\frac{{\dot{b}}^{2}}{6{b}^{2}}+\frac{16{\dot{b}}^{3}\dddot{b}\beta}{\Mpl^{4}{b}^{4}}+\frac{8{\ddot{b}}^{2}{\dot{b}}^{2}\beta}{\Mpl^{4}{b}^{4}}-\frac{32\ddot{b}{\dot{b}}^{4}\beta}{\Mpl^{4}{b}^{5}}+\frac{\dot{b}\dot{a}}{3ba}-\frac{8{\dot{b}}^{3}\dddot{a}\beta}{\Mpl^{4}{b}^{3}a}-\frac{24{\dot{b}}^{2}\dddot{b}\dot{a}\beta}{\Mpl^{4}{b}^{3}a}-\frac{8\ddot{b}{\dot{b}}^{2}\ddot{a}\beta}{\Mpl^{4}{b}^{3}a}-\frac{8{\ddot{b}}^{2}\dot{b}\dot{a}\beta}{\Mpl^{4}{b}^{3}a}\nonumber \\
 &  & {}+\frac{48\ddot{b}{\dot{b}}^{3}\dot{a}\beta}{\Mpl^{4}{b}^{4}a}+\frac{16{\dot{b}}^{5}\dot{a}\beta}{\Mpl^{4}{b}^{5}a}+\frac{8\beta\,{\dot{b}}^{2}\dot{a}\dddot{a}}{\Mpl^{4}{b}^{2}{a}^{2}}+\frac{8\beta\,{\dot{b}}^{2}{\ddot{a}}^{2}}{\Mpl^{4}{b}^{2}{a}^{2}}+\frac{8\beta\,{\dot{a}}^{2}\dot{b}\dddot{b}}{\Mpl^{4}{b}^{2}{a}^{2}}-\frac{8\beta\,\dot{a}\dot{b}\ddot{b}\ddot{a}}{\Mpl^{4}{b}^{2}{a}^{2}}+\frac{8\beta\,{\ddot{b}}^{2}{\dot{a}}^{2}}{\Mpl^{4}{b}^{2}{a}^{2}}+\frac{8\beta\,\dot{a}{\dot{b}}^{3}\ddot{a}}{\Mpl^{4}{b}^{3}{a}^{2}}\nonumber \\
 &  & {}-\frac{8\beta\,{\dot{a}}^{2}{\dot{b}}^{2}\ddot{b}}{\Mpl^{4}{b}^{3}{a}^{2}}-\frac{24\beta\,{\dot{a}}^{2}{\dot{b}}^{4}}{\Mpl^{4}{b}^{4}{a}^{2}}-\frac{8\beta\,{\dot{a}}^{2}{\dot{b}}^{2}\ddot{a}}{\Mpl^{4}{b}^{2}{a}^{3}}-\frac{8{\dot{a}}^{3}\beta\,\dot{b}\ddot{b}}{\Mpl^{4}{b}^{2}{a}^{3}}+\frac{16{\dot{a}}^{3}\beta\,{\dot{b}}^{3}}{\Mpl^{4}{b}^{3}{a}^{3}}=0\,,\\
E_{2} & \equiv & \frac{\ddot{b}}{3b}+\frac{8\dot{b}\ddddot{b}\dot{a}\beta}{\Mpl^{4}{b}^{2}a}+\frac{8\dot{b}\dddot{b}\ddot{a}\beta}{\Mpl^{4}{b}^{2}a}+\frac{24\ddot{b}\dddot{b}\dot{a}\beta}{\Mpl^{4}{b}^{2}a}-\frac{8{\dot{b}}^{2}\ddddot{b}\beta}{\Mpl^{4}{b}^{3}}-\frac{32\ddot{b}\dot{b}\dddot{b}\beta}{\Mpl^{4}{b}^{3}}+\frac{8{\ddot{b}}^{2}\ddot{a}\beta}{\Mpl^{4}{b}^{2}a}-\frac{8{\dot{b}}^{4}\ddot{a}\beta}{\Mpl^{4}{b}^{4}a}-\frac{16{\dot{a}}^{3}\beta\,{\dot{b}}^{3}}{\Mpl^{4}{b}^{3}{a}^{3}}\nonumber \\
 &  & {}+\frac{16{\dot{b}}^{5}\dot{a}\beta}{\Mpl^{4}{b}^{5}a}-\frac{16\beta\,{\ddot{b}}^{2}{\dot{a}}^{2}}{\Mpl^{4}{b}^{2}{a}^{2}}-\frac{8\beta\,{\dot{a}}^{2}{\dot{b}}^{4}}{\Mpl^{4}{b}^{4}{a}^{2}}-\frac{8{\ddot{b}}^{3}\beta}{\Mpl^{4}{b}^{3}}-\frac{8{\dot{b}}^{2}\dddot{b}\dot{a}\beta}{\Mpl^{4}{b}^{3}a}-\frac{8\ddot{b}{\dot{b}}^{2}\ddot{a}\beta}{\Mpl^{4}{b}^{3}a}-\frac{16{\ddot{b}}^{2}\dot{b}\dot{a}\beta}{\Mpl^{4}{b}^{3}a}\nonumber \\
 &  & {}-\frac{24\ddot{b}{\dot{b}}^{3}\dot{a}\beta}{\Mpl^{4}{b}^{4}a}-\frac{16\beta\,{\dot{a}}^{2}\dot{b}\dddot{b}}{\Mpl^{4}{b}^{2}{a}^{2}}+\frac{24\beta\,\dot{a}{\dot{b}}^{3}\ddot{a}}{\Mpl^{4}{b}^{3}{a}^{2}}+\frac{40\beta\,{\dot{a}}^{2}{\dot{b}}^{2}\ddot{b}}{\Mpl^{4}{b}^{3}{a}^{2}}+\frac{16{\dot{a}}^{3}\beta\,\dot{b}\ddot{b}}{\Mpl^{4}{b}^{2}{a}^{3}}+\frac{24{\dot{b}}^{3}\dddot{b}\beta}{\Mpl^{4}{b}^{4}}\nonumber \\
 &  & {}+\frac{64{\ddot{b}}^{2}{\dot{b}}^{2}\beta}{\Mpl^{4}{b}^{4}}-\frac{32\ddot{b}{\dot{b}}^{4}\beta}{\Mpl^{4}{b}^{5}}+\frac{{\dot{b}}^{2}}{6{b}^{2}}-\frac{24\beta\,\dot{a}\dot{b}\ddot{b}\ddot{a}}{\Mpl^{4}{b}^{2}{a}^{2}}-\frac{\Lambda}{6}=0\,,\\
E_{3} & \equiv & -\frac{24\dot{a}\beta\,\dot{b}{\ddot{a}}^{2}}{\Mpl^{4}b{a}^{3}}-\frac{32\ddot{b}\dot{b}\dddot{a}\beta}{\Mpl^{4}{b}^{2}a}+\frac{24\beta\,\dot{b}\ddot{a}\dddot{a}}{\Mpl^{4}b{a}^{2}}-\frac{8{\dot{a}}^{2}\beta\,\ddot{a}\ddot{b}}{\Mpl^{4}b{a}^{3}}+\frac{8\beta\,\ddot{b}\dot{a}\dddot{a}}{\Mpl^{4}b{a}^{2}}+\frac{16\dot{b}\ddot{a}{\dot{a}}^{3}\beta}{\Mpl^{4}b{a}^{4}}-\frac{48{\dot{b}}^{5}\dot{a}\beta}{\Mpl^{4}{b}^{5}a}+\frac{8\beta\,{\dot{b}}^{2}{\ddot{a}}^{2}}{\Mpl^{4}{b}^{2}{a}^{2}}\nonumber \\
 &  & {}+\frac{8\dot{a}\beta\,\dot{b}\ddddot{a}}{\Mpl^{4}b{a}^{2}}-\frac{16{\dot{a}}^{2}\beta\,\dot{b}\dddot{a}}{\Mpl^{4}b{a}^{3}}+\frac{\ddot{a}}{3a}-\frac{16{\dot{b}}^{2}{\dot{a}}^{4}\beta}{\Mpl^{4}{b}^{2}{a}^{4}}+\frac{8\beta\,\ddot{b}{\ddot{a}}^{2}}{\Mpl^{4}b{a}^{2}}-\frac{8{\dot{b}}^{2}\ddddot{a}\beta}{\Mpl^{4}{b}^{2}a}+\frac{64\ddot{b}\dot{b}\dddot{b}\beta}{\Mpl^{4}{b}^{3}}\nonumber \\
 &  & {}-\frac{24{\ddot{b}}^{2}\ddot{a}\beta}{\Mpl^{4}{b}^{2}a}+\frac{16{\dot{b}}^{4}\ddot{a}\beta}{\Mpl^{4}{b}^{4}a}+\frac{\ddot{b}}{3b}-\frac{32\ddot{b}\dddot{b}\dot{a}\beta}{\Mpl^{4}{b}^{2}a}-\frac{16\dot{b}\ddddot{b}\dot{a}\beta}{\Mpl^{4}{b}^{2}a}-\frac{32\dot{b}\dddot{b}\ddot{a}\beta}{\Mpl^{4}{b}^{2}a}+\frac{16\dot{b}^{2}\ddddot{b}\beta}{\Mpl^{4}{b}^{3}}\nonumber \\
 &  & {}-\frac{\Lambda}{3}-\frac{64{\dot{b}}^{3}\dddot{b}\beta}{\Mpl^{4}{b}^{4}}-\frac{144{\ddot{b}}^{2}{\dot{b}}^{2}\beta}{\Mpl^{4}{b}^{4}}+\frac{96\ddot{b}{\dot{b}}^{4}\beta}{\Mpl^{4}{b}^{5}}+\frac{16\dot{a}\beta\,\dot{b}\ddot{b}\ddot{a}}{\Mpl^{4}{b}^{2}{a}^{2}}+\frac{16{\ddot{b}}^{3}\beta}{\Mpl^{4}{b}^{3}}+\frac{\dot{b}\dot{a}}{3ba}\nonumber \\
 &  & {}+\frac{64{\dot{b}}^{2}\dddot{b}\dot{a}\beta}{\Mpl^{4}{b}^{3}a}+\frac{56\ddot{b}{\dot{b}}^{2}\ddot{a}\beta}{\Mpl^{4}{b}^{3}a}+\frac{96{\ddot{b}}^{2}\dot{b}\dot{a}\beta}{\Mpl^{4}{b}^{3}a}-\frac{16\ddot{b}{\dot{b}}^{3}\dot{a}\beta}{\Mpl^{4}{b}^{4}a}+\frac{8\beta\,{\dot{b}}^{2}\dot{a}\dddot{a}}{\Mpl^{4}{b}^{2}{a}^{2}}+\frac{32{\dot{a}}^{2}\beta\,{\dot{b}}^{4}}{\Mpl^{4}{b}^{4}{a}^{2}}+\frac{16{\dot{a}}^{3}\beta\,{\dot{b}}^{3}}{\Mpl^{4}{b}^{3}{a}^{3}}\nonumber \\
 &  & {}-\frac{48\dot{a}\beta\,{\dot{b}}^{3}\ddot{a}}{\Mpl^{4}{b}^{3}{a}^{2}}-\frac{64{\dot{a}}^{2}\beta\,{\dot{b}}^{2}\ddot{b}}{\Mpl^{4}{b}^{3}{a}^{2}}+\frac{16{\dot{a}}^{2}\beta\,{\dot{b}}^{2}\ddot{a}}{\Mpl^{4}{b}^{2}{a}^{3}}+\frac{16{\dot{a}}^{3}\beta\,\dot{b}\ddot{b}}{\Mpl^{4}{b}^{2}{a}^{3}}+\frac{8{\dot{b}}^{3}\dddot{a}\beta}{\Mpl^{4}{b}^{3}a}=0\,.
\end{eqnarray}
As long as we are not in the exact FLRW limit, then we can solve these
equations for $\Lambda$, $\ddddot{a}$ and $\ddddot{b}$. These equations
in the text will be denoted as $\Lambda$, $a$ and $b$ equations
respectively. In the process of reducing the second order Lagrangian density
for the perturbations we will make use of these equations of motion.
It should be noted that $E_{1}$, $E_{2}$, and $E_{3}$ are not independent
equations of motion, in fact we have the following identity 
\begin{equation}
\dot{E}_{1}+\left(\frac{\dot{a}}{a}+\frac{2\dot{b}}{b}\right)E_{1}-\frac{\dot{a}}{a}\,E_{2}-\frac{\dot{b}}{b}\,E_{3}=0\,.
\end{equation}
This relation states that $E_{3}$ can be written in terms of $E_{1}$,
its time derivative and $E_{2}$. Therefore, once both $E_{1}$ and
$E_{2}$ are satisfied, then automatically also $E_{3}$ will be.

\section{Regular field redefinition\label{sec:Regular-field-redefinition}}

We will write here the field redefinition which diagonalizes the kinetic
matrix for a general Bianchi-I solution for the theory under consideration.
We write it explicitly here to show it is in general regular, in particular
in the isotropic limit. We have in general that $K_{23}\propto(a\dot{b}-b\dot{a})^{2}$,
whereas both $K_{12}$ and $K_{33}$ are proportional to $(a\dot{b}-b\dot{a})$
linearly. Then we obtain, without any approximation, i.e.\ for a
general Bianchi-I manifold 
\begin{eqnarray}
\Gamma_{1} & = & -\frac{K_{12}K_{33}}{K_{22}K_{33}-K_{23}^{2}}=-\frac{24(\dot{b}a-\dot{a}b)\,\dot{b}\beta\,q^{2}b^{2}a^{4}}{\Delta_{1}}\,,\\
\Gamma_{2} & = & \frac{K_{23}K_{12}}{K_{22}K_{33}-K_{23}^{2}}=-\frac{24(\dot{b}a-\dot{a}b)^{2}\,\dot{b}\beta\,k^{2}b^{3}a^{2}}{\Delta_{1}}\,,\\
\Gamma_{3} & = & -\frac{K_{23}}{K_{33}}=\frac{k^{2}b}{q^{2}a^{2}}\,(a\dot{b}-\dot{a}b)\,,\\
\Delta_{1} & = & -q^{2}a^{5}\Mpl^{4}{b}^{4}+\beta\left\{ 48\,{q}^{2}\left[\left({\ddot{b}}^{2}+\frac{3}{2}\,\dddot{b}\dot{b}\right){b}^{2}+\frac{1}{2}\,\ddot{b}\left({q}^{2}-13\,{\dot{b}}^{2}\right)b+\frac{1}{2}\,{q}^{2}{\dot{b}}^{2}+{\dot{b}}^{4}\right]{a}^{5}\right.\nonumber \\
 &  & {}+24\,b{q}^{2}\left[\ddot{a}\ddot{b}{b}^{2}+\left(\ddot{a}{q}^{2}+3\,\dot{a}\ddot{b}\dot{b}+2\,\ddot{a}{\dot{b}}^{2}\right)b-3\,{q}^{2}\dot{a}\dot{b}+5\,\dot{a}{\dot{b}}^{3}\right]{a}^{4}\nonumber \\
 &  & {}-24\,{b}^{2}\left[\left(3\,\dot{a}\ddot{a}\dot{b}+\ddot{b}({k}^{2}+{\dot{a}}^{2})\right)q^{2}b-\left(q^{2}(k^{2}-7\,\dot{a}^{2})-2\,\dot{b}^{2}k^{2}\right){\dot{b}}^{2}\right]{a}^{3}\nonumber \\
 &  & {}+\left.144\,{b}^{3}\dot{b}\dot{a}\left({\dot{b}}^{2}{k}^{2}+\frac{2}{3}\,{q}^{2}{\dot{a}}^{2}\right){a}^{2}-144\,a{b}^{4}{k}^{2}{\dot{a}}^{2}{\dot{b}}^{2}+48\,{b}^{5}{k}^{2}{\dot{a}}^{3}\dot{b}\right\} ,
\end{eqnarray}
and we can see that in the exact isotropic limit, we find that $\Gamma_{1}$,
$\Gamma_{2}$, and $\Gamma_{3}$ all vanish as 
\begin{equation}
\lim_{a,b\to a_{0}}\Delta_{1}=-q^{2}a_{0}^{9}\,(\Mpl^{4}+48\,\beta\,H_{0}^{4})\,.
\end{equation}
Then, after performing the field redefinition, we find 
\begin{eqnarray}
g_{1} & = & \frac{144\,q^{4}\dot{b}^{2}\beta^{2}k^{2}\,(a\dot{b}-b\dot{a})^{2}a^{4}}{\Mpl^{2}\Delta_{1}}\,,\\
g_{2} & = & -\frac{k^{2}\Delta_{1}}{4a^{4}b^{4}\Mpl^{2}}\,,\\
g_{3} & = & -\frac{12\,(a\dot{b}-b\dot{a})\,\dot{b}q^{4}\beta}{\Mpl^{2}b^{4}}\,.
\end{eqnarray}

\subsection{Exact FLRW limit}

We discuss here the exact limit for which $b\to a$ in the Lagrangian density
for the fields $F_{i}$. Although this limit should \emph{not} be
made exactly, as we would loose information regarding the propagating
fields $F_{1}$ and $F_{2}$, we only want to show here that we can
get back the FLRW result for the cross polarization of gravitational waves. In 
fact, in this case the Lagrangian density reduces to 
\begin{eqnarray}
\mathcal{L} & = & \frac{\Mpl^{4}+48\,\beta\,H_{0}^{4}}{\Mpl^{2}}\left[\frac{a_{0}q^{2}k^{2}}{4}\,\dot{F}_{2}^{2}-\frac{q^{2}k^{2}}{4}\,(\dot{F}_{2}F_{3}-\dot{F}_{3}F_{2})-\frac{(q^{2}-2\,H_{0}^{2}\,a_{0}^{2})q^{2}k^{2}}{4a_{0}}\,F_{2}^{2}\right.\nonumber \\
 &  & {}+\left.\frac{q^{2}k^{2}H_{0}}{2}\,F_{2}F_{3}+\frac{(k^{2}+q^{2})\,q^{2}}{4a_{0}}\,F_{3}^{2}\right],
\end{eqnarray}
and any term including the field $F_{1}$ disappears in this exact
FLRW limit. After integrating by parts the term $F_{2}\dot{F}_{3}$
term, we find that the field $F_{3}$ becomes a Lagrange multiplier,
which can be integrated out (at least in this \emph{wrong} limit)
to give 
\begin{equation}
F_{3}=-\frac{a_{0}k^{2}\,(H_{0}F_{2}-\dot{F}_{2})}{k^{2}+q^{2}}\,.
\end{equation}
On performing the field redefinition 
\begin{equation}
F_{2}=\frac{\sqrt{2}\,a_{0}\sqrt{k^{2}+q^{2}}}{2kq^{2}}\,f_{2}\,,
\end{equation}
then we get, as expected, the standard propagation for a gravitational wave in the de
Sitter background for this theory, namely 
\begin{equation}
\mathcal{L}=\frac{(\Mpl^{4}+48\,\beta\,H_{0}^{4})\,a_{0}^{3}}{8\Mpl^{2}}\left[\dot{f}_{2}^{2}-\frac{k^{2}+q^{2}}{a_{0}^{2}}\,f_{2}^{2}\right].
\end{equation}

\acknowledgments

A.\ D.\ F.\ was supported by JSPS KAKENHI Grant Number 20K03969. C.\ P.\ M.\ acknowledges the support from the Japanese Government
(MEXT) scholarship for Research Student. A.\ A.\ S.\  is supported by the RSF grant 16-12-10401.

 \bibliographystyle{unsrt}
\bibliography{bibliography}

\end{document}